\documentclass[letterpaper,aps,prd,twocolumn,floats, floatfix,showpacs]{revtex4-1}
\usepackage{graphicx}
\usepackage{amsmath,amssymb,amsfonts}

\begin{document}

\title{Fermi gamma-ray `bubbles' from stochastic acceleration of electrons} 
\author{Philipp Mertsch}
\affiliation{Rudolf Peierls Centre for Theoretical Physics, University
  of Oxford, Oxford OX1 3NP, UK}
\author{Subir Sarkar}
\affiliation{Rudolf Peierls Centre for Theoretical Physics, University
  of Oxford, Oxford OX1 3NP, UK}
\begin{abstract}
  Gamma-ray data from Fermi-LAT reveal a bi-lobular structure
  extending up to $\sim50{}^\circ$ above and below the galactic
  centre, which presumably originated in some form of energy release
  there less than a few million years ago. It has been argued that the
  $\gamma$-rays arise from hadronic interactions of high energy cosmic
  rays which are advected out by a strong wind, or from
  inverse-Compton scattering of relativistic electrons accelerated at
  plasma shocks present in the bubbles. We explore the alternative
  possibility that the relativistic electrons are undergoing
  stochastic 2$^\text{nd}$-order Fermi acceleration by plasma wave
  turbulence through the entire volume of the bubbles. The observed
  $\gamma$-ray spectral shape is then explained naturally by the
  resulting hard electron spectrum modulated by inverse-Compton energy
  losses. Rather than a constant volume emissivity as in other models,
  we predict a nearly constant surface brightness, and reproduce the
  observed sharp edges of the bubbles.
\end{abstract}
\date{\today}
\maketitle

Recent data from the Fermi-LAT satellite has
revealed~\cite{Su:2010qj,Fermi} the presence of giant $\gamma$-ray
lobes $\sim40{}^\circ$ wide, extending up to $\sim50{}^\circ$ above
and below the galactic centre (GC). The energy spectrum of the
emission from these `Fermi bubbles' is $\text{d}N/\text{d}E
\sim\!E^{-2}$ from $\sim\!1-200\,\text{GeV}$, \emph{i.e.}
considerably harder than conventional foregrounds. Furthermore, the
bubbles exhibit an almost constant surface brightness with hard edges.
While the template subtraction technique used to reveal the bubbles
may not be appropriate at these high energies, the resulting
systematic effects are not easy to assess.  However the bubbles do
correlate with features at other wavelengths, \emph{viz.} data from
the ROSAT X-ray satellite~\cite{Snowden:1997ze} show a
limb-brightened, conical structure close to the galactic plane which
coincides with the edges of the Fermi bubbles. The bubbles also line
up with a claimed excess in microwaves at lower galactic latitudes ---
the so-called `WMAP haze'~\cite{Finkbeiner:2004us}.

Although extended lobes have long been seen in other galaxies in
radio, X-rays and $\gamma$-rays, their presence in the Milky Way is
surprising. There is no radio emission from these bubbles, unlike
those seen in the majority of active galaxies. Moreover their
morphology (symmetry with respect to the galactic plane and alignment
with the GC) suggests that the central supermassive black hole is the
energy source. However it is supposedly in a quiescent state so it is
a puzzle how the bubbles have formed; understanding this would provide
an excellent probe of this region which is otherwise obscured by the
galactic disk.  The bubbles may play an important role in the dynamics
of our galaxy and constitute a source of cosmic rays (CR). While they
are prominent at high galactic latitudes, the associated signal close
to the plane, while uncertain, constitutes a background for indirect
dark matter searches. It is therefore important to understand and
model the origin of the non-thermal emission from the bubbles.

While the mechanism responsible for the formation of the bubbles is
not necessarily the same as the source of the $\gamma$-ray emission
today, it is useful to recall their general properties. The
limb-brightened shell in the ROSAT data implies a shock front at the
bubble edges, but from the observed cavity hot low density gas is
inferred to fill the bubble interiors. Assuming a low density ($n \sim
10^{-2}\,\text{cm}^{-3}$) gas at $T \sim 2\,\text{keV}$ and shock
velocities $U \lesssim 1000\,\text{km}\,\text{s}^{-1}$, the energy is
estimated to be $\sim\!10^{54-55}\,\text{erg}$ in hot gas and the age
to be $\sim \!10^7 (U/1000\,\text{km} \,\text{s}^{-1})
\text{yr}$~\cite{Su:2010qj}. Suggested mechanisms for providing such
an energy on this timescale include jets emanating from the central
black hole~\cite{Guo:2011eg}, star forming regions close to the
GC~\cite{Crocker:2010dg} or repeated star accretion onto the central
black hole~\cite{Cheng:2011xd}.

The observed $\gamma$-rays may be generated by hadronic interactions
of high-energy CR protons or nuclei (\emph{i.e.} $\pi^0$ decay)
provided that the ambient gas-density is not too low. It has been
proposed~\cite{Crocker:2010dg} that protons and nuclei accelerated by
supernova remnants (SNRs) in star-forming regions very close to the GC
could be advected by a strong wind out to kiloparsec distances above
the plane. If the confinement time is larger than all other
timescales, the hard power law spectrum of the $\gamma$-rays would
simply reflect the source spectrum of the protons. The spectral
shoulder at $\sim\!1\,\text{GeV}$ can be explained by the pion bump.

Another possible mechanism is the inverse-Compton (IC) scattering of
high energy electrons off ambient radiation fields (CMB, far infra-red
(FIR) and optical/UV). The spectral feature seen at a few hundred GeV
may reflect a cut-off in the electron spectrum at a similar energy,
either due to energy losses or due to the competition between an
energy-dependent acceleration rate and the finite age of the
bubbles. Furthermore, the WMAP haze may well arise from synchrotron
radiation of these electrons in the ambient magnetic field. A crucial
question then is how are the electrons accelerated.

The standard paradigm for the acceleration of galactic CRs is
diffusive shock acceleration (DSA) by the 1$^\text{st}$-order Fermi
process, which predicts power-law source spectra with index close to
$-2$. There are at least four regions where shocks may be present: at
the GC, inside a jet emanating from the GC, at its termination shock
at the upper/lower edges of the bubbles, and at the shocked exterior
of the bubbles. So far there is only evidence from ROSAT data for a
shock at the bubble exterior.  In any case, presuming
diffusive-convective transport from the acceleration site through the
bubble volume, it is difficult to see how the electrons can maintain
their hard source spectrum. The energy loss time due to IC scattering
for the $\mathcal{O}(\text{TeV})$ energy electrons present throughout
the bubble is only a few times $10^5 \,\text{yr}$; however even with a
convection velocity as high as $v \sim
1000\,\text{km}\,\text{s}^{-1}$, it would take the electrons
$10^7\,\text{yr}$ to cross the required distance of
$\mathcal{O}(10)\,\text{kpc}$. The leptonic source model
\cite{Cheng:2011xd} therefore invokes \emph{hundreds of consecutive
  shocks} in order to fill the whole bubble with freshly accelerated
electrons. This would however imply a constant volume emissivity which
in projection would yield a characteristic bump-like profile with soft
edges, in contrast to what is observed.

We consider instead the stochastic acceleration of high energy
electrons by isotropic, large-scale turbulence in magnetosonic
waves~\cite{Ptuskin:1988aa}. Such 2$^\text{nd}$-order Fermi
acceleration accounts well for the radio emission from supernova
remnants~\cite{Scott:1975aa,Fan:2009kr} and the extended lobes of
radio galaxies~\cite{Lacombe:1977aa}, and may even be the acceleration
mechanism for ultra-high energy cosmic rays~\cite{Hardcastle:2008jw}.
The shock front at the bubble edges suggests that they may have been
powered by a jet emanating from the massive black hole at the GC that
was active a few million years ago. MHD modelling~\cite{Guo:2011eg} of
a two-component plasma explains the formation of a bubble by a light
but over-pressured jet with $\sim\!16\,\%$ of the Eddington luminosity,
and also predicts a shock coincident with the ROSAT shell. Plasma
instabilities, in particular Rayleigh-Taylor and Kelvin-Helmholtz
instabilities, would then generate turbulence at the outer shock that
is convected into the bubble interior by the downstream plasma flow.
The free energy dissipation rate $Q = C_1 \rho u^3 / L$ is determined
by the scale of turbulence injection, $L$, and the eddy velocity at
the injection scale, $u=v_{\text{edd}}(L)$, where $C_1 = 0.485$ is the
1-dimensional Kolmogorov constant~\cite{Fan:2009kr}.  The energy
density at scale $k$ is then given by $W(k) = (u^2 / 4\pi) L^{-2/3}
k^{-11/3}$.  Applying the Rankine-Hugoniot conditions at the shock,
the eddy velocity at the injection scale, $u$, and the magnetosonic
phase velocity, $v_\text{F}$, vary with the distance $x = \xi L$ from the
shock as~\cite{Fan:2009kr}:
\begin{align}
u(\xi) &= \frac{U}{4} \frac{1}{C_1 \xi/3 + a^{-1/2}}\,, \label{eqn:u} \\
v_\text{F}(\xi) &= \frac{U}{4} \left( {5 - \frac{5}{3(C_1 \xi/3)^2} 
 + 4 \frac{v_\text{A}^2}{U^2}} \right)^{1/2} \,, \label{eqn:vF}
\end{align}
where $U$ is the shock velocity, $v_\text{A}$ the Alfv\'en velocity
(which we assume to be constant and equal to the speed of sound
$v_{\text{s}, 0}$ at the shock) and $a = 3 - 16 v_{\text{s}, 0}^2 /
U^2$.

At small enough scales $l_\text{d} = 1/k_\text{d} = L
(v_\text{A}/u)^3$, the kinetic energy of the turbulence becomes
comparable to the magnetic field energy, $v_{\text{edd}}(l_\text{d})
\approx v_\text{A}$, resulting in transit-time damping. With
parameters to be justified below, it turns out that for all energies
of interest, the gyro-radius $r_\text{g}$ of the electrons is always
smaller than this dissipation scale $l_\text{d}$ such that
gyro-resonant interactions with magnetosonic turbulence are not
possible. Therefore, we adopt the dissipation scale to be the
mean-free path, thus rendering the spatial diffusion coefficient
$D_{xx} = l_\text{d} c/3$ energy independent. If additional
\textit{small-scale} turbulence is present (possibly responsible for
spatial diffusion~\cite{Ptuskin:1988aa}), then the mean-free path can
be smaller.

The temporal evolution of $n(t,p) \, {\rm d}p$, the number density of
electrons with momentum between $p$ and $(p+{\rm d}p)$, is dictated by
the Fokker-Planck equation~\cite{Ginzburg:1957aa},
\begin{equation}
 \frac{\partial n}{\partial t} - \frac{\partial}{\partial p} 
 \left( p^2 D_{pp} \frac{\partial}{\partial p} \frac{n}{p^2} \right)  
 - \frac{n}{t_{\text{esc}}} + \frac{\partial}{\partial p} 
 \left( \frac{\text{d}p}{\text{d}t} n \right) = 0 \,,
\label{eqn:FokkerPlanck}
\end{equation}
where the diffusion coefficient in momentum for scattering by fast
magnetosonic waves is ~\cite{Ptuskin:1988aa}:
\begin{equation}
D_{pp} = p^2 \frac{8 \pi D_{xx}}{9} \int_{1/L}^{k_\text{d}}\text{d} k \, 
 \frac{W(k) k^4}{v_\text{F}^2 + D_{xx}^2 k^2}\, .
\end{equation}
The second term in Eq.~\ref{eqn:FokkerPlanck} describes diffusion in
momentum as well as systematic energy gains on the characteristic
timescale $t_{\text{acc}} \sim p^2 / D_{pp}$, which is also energy
independent. Diffusive losses from the acceleration region can be
accounted for by escape on the timescale $t_{\text{esc}} =
L^2/D_{xx}$. Finally, the electrons lose energy through IC scattering
and synchrotron radiation which are both accounted for by the energy
dependent cooling time \mbox{$t_{\text{cool}} =
  -p/(\text{d}p/\text{d}t)$}.
\begin{figure}[!bh]
\begin{center}
\includegraphics[width=\linewidth]{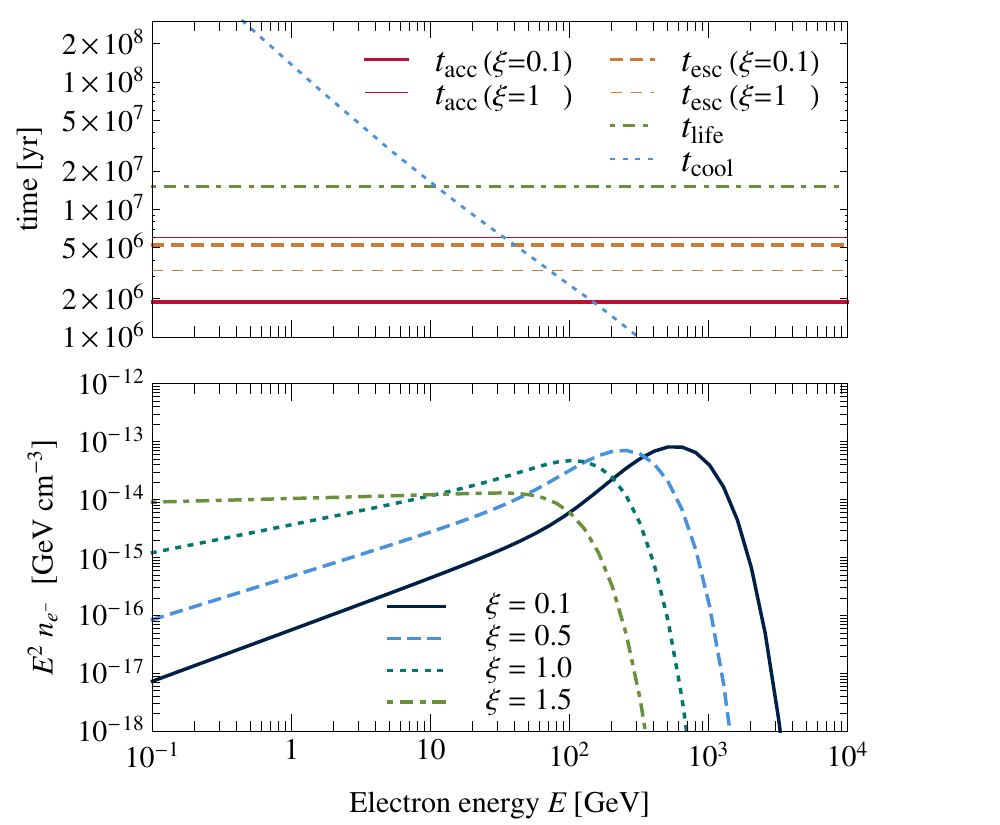}
\end{center}
\caption[]{Relevant timescales (top) and the electron spectrum
  (bottom), at various distances $x = \xi L$ from the shock.}
\label{fig1}
\end{figure}

Because of the energy-independent spatial diffusion coefficient, the
so-called ``hard-sphere'' approximation~\cite{Kardashev:1962aa} is
\emph{exact} which makes the problem amenable to analytical
solution. If the escape rate is not much bigger than the acceleration
rate, \emph{i.e.} $t_{\text{acc}} \lesssim t_{\text{esc}}$, the steady
state spectrum $n(p)$ at a fixed position can be described as a power
law with a spectral cut-off above (and pile-up around) a
characteristic momentum $p_{\text{eq}}$, defined by
$t_{\text{acc}}(p_{\text{eq}}) \equiv
t_{\text{cool}}(p_{\text{eq}})$~\cite{Stawarz:2008sp}:
\begin{equation}
n(p) \propto \left \{
\begin{array}{ll}
p^{-\sigma} & \quad \text{for} \quad p \ll p_{\text{eq}} \, , \\
p^2 {\rm e}^{-p/p_{\text{eq}}} & \quad \text{for} \quad p \sim p_{\text{eq}} \,.
\end{array}
\right.
\end{equation}
The spectral index, $-\sigma = 1/2 - \sqrt{9/4 +
  t_\text{acc}/t_\text{esc}}$, is determined by the ratio of
acceleration and escape times, and asymptotically approaches $-1$
as $t_\text{acc}/t_\text{esc} \rightarrow 0$.

Anticipating that the acceleration time is smaller than the lifetime
$t_{\text{life}}$ of the bubbles, we justify the use of the
steady-state solution for acceleration volumes that are being advected
with the downstream plasma. It is sufficient to consider the variation
of the acceleration and escape times with the distance from the shock,
which determines the spatial dependence of the electron spectrum. This
hierarchy of timescales assures that the variation with position
happens \emph{adiabatically}, such that the electrons can always relax
to their steady-state spectrum.  In the upper panel of Fig.~\ref{fig1}
we show the different timescales in the problem as a function of
energy for the parameters discussed below. Although $t_{\text{cool}}$
is of the same order as the dynamical time $t_{\text{conv}}$ around
$10 \, \text{GeV}$, we expect that the steady state spectrum is
reached in a time $t \sim t_{\text{acc}}$, as has been shown
explicitly~\cite{Becker:2006nz} for ionisation losses.

The relative normalisation of the electron spectrum is fixed by noting
that the total energy in relativistic electrons at any position is a
constant fraction of the free energy dissipated along with the
downstream plasma up until this position. This does not however fix
the absolute normalisation which depends on the microphysics of the
acceleration process, in particular the injection mechanism. We
determine the $\gamma$-ray volume emissivity due to IC scattering off
the CMB, FIR and optical/UV backgrounds adopting the interstellar
radiation fields from \texttt{GALPROP}~\cite{Porter:2005qx} at a
reference height of $4\,\text{kpc}$ above the GC.  For the parameters
discussed below we show the electron spectrum $E^2 n_{e^{-}}$ for
different distances from the shock in the lower panel of
Fig.~\ref{fig1}.

\begin{figure}[!t]
\begin{center}
\includegraphics[width=\linewidth]{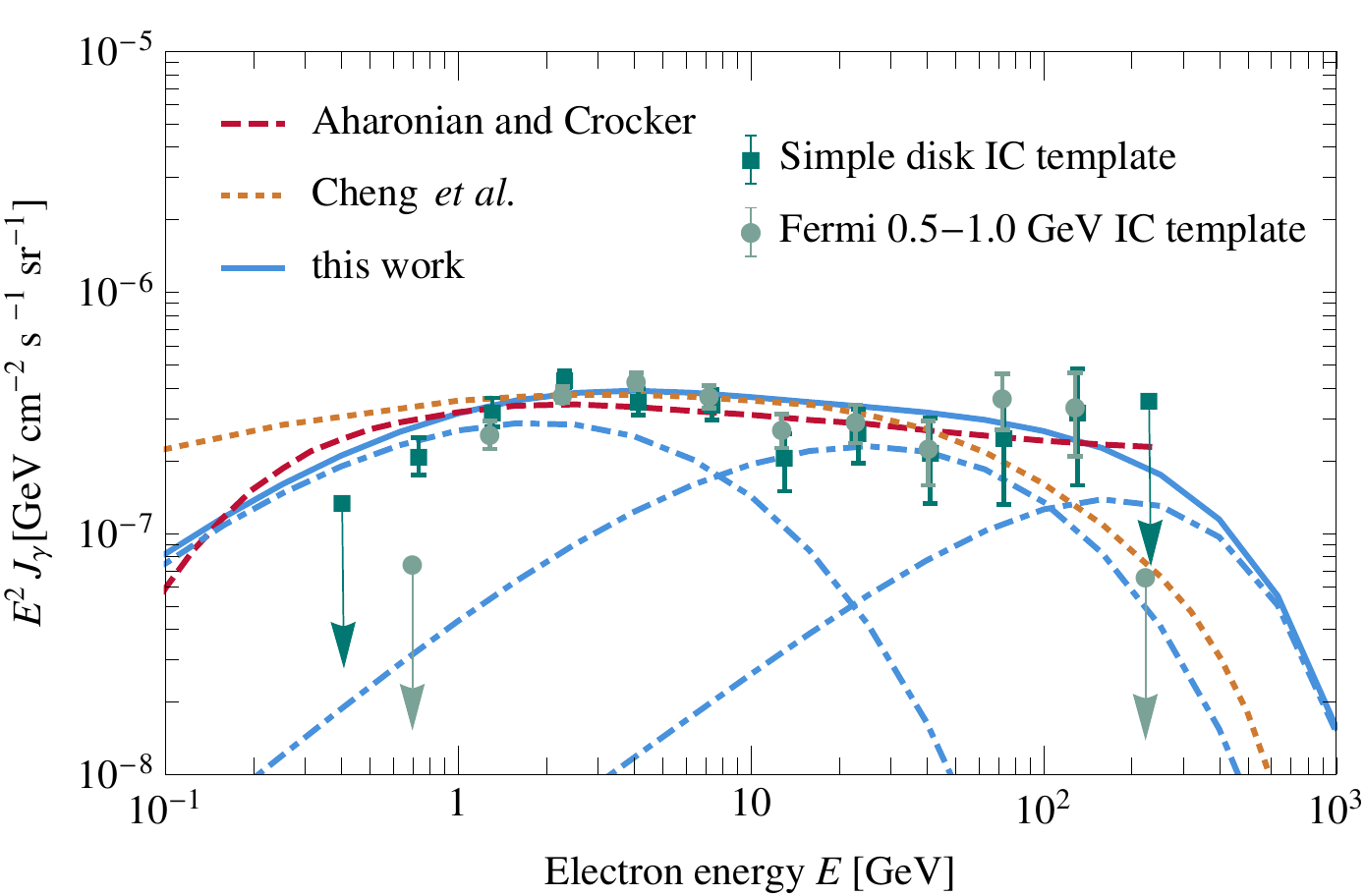}
\end{center}
\caption[]{The average $\gamma$-ray flux $J_{\gamma}$ from the Fermi
  bubbles~\cite{Su:2010qj} compared to the spectrum from our model
  (the contributions from inverse-Compton scattering on the CMB, FIR
  and optical/UV backgrounds are shown from left to right as
  dot-dashed lines). The spectra from a hadronic~\cite{Crocker:2010dg}
  (dashed line) and a leptonic DSA model~\cite{Cheng:2011xd} (dotted
  line) are also shown.}
\label{fig2}
\end{figure}

\begin{figure}[!b]
\begin{center}
\includegraphics[width=\linewidth]{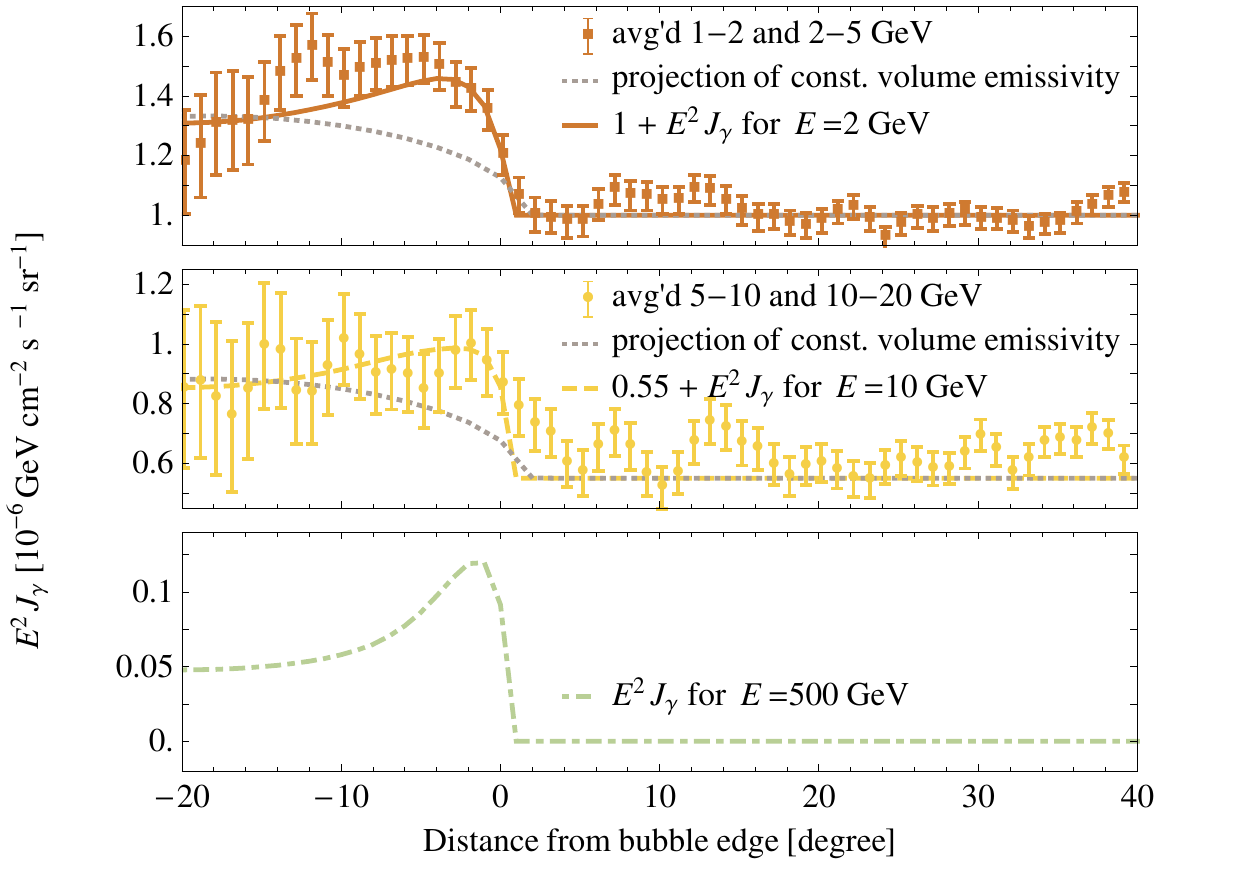}
\end{center}
\caption[]{The $\gamma$-ray intensity as a function of distance from
  the bubble edge~\cite{Su:2010qj} is compared with our model
  predictions at $2\,\text{GeV}$ (solid), $10\,\text{GeV}$ (long
  dashed), and $500\,\text{GeV}$ (dot-dashed). The dotted line
  indicates the expected profile for both the hadronic
  model~\cite{Crocker:2010dg} and the leptonic DSA
  model~\cite{Cheng:2011xd}.}
\label{fig3}
\end{figure}

\begin{figure}[!t]
\begin{center}
\includegraphics[width=\linewidth]{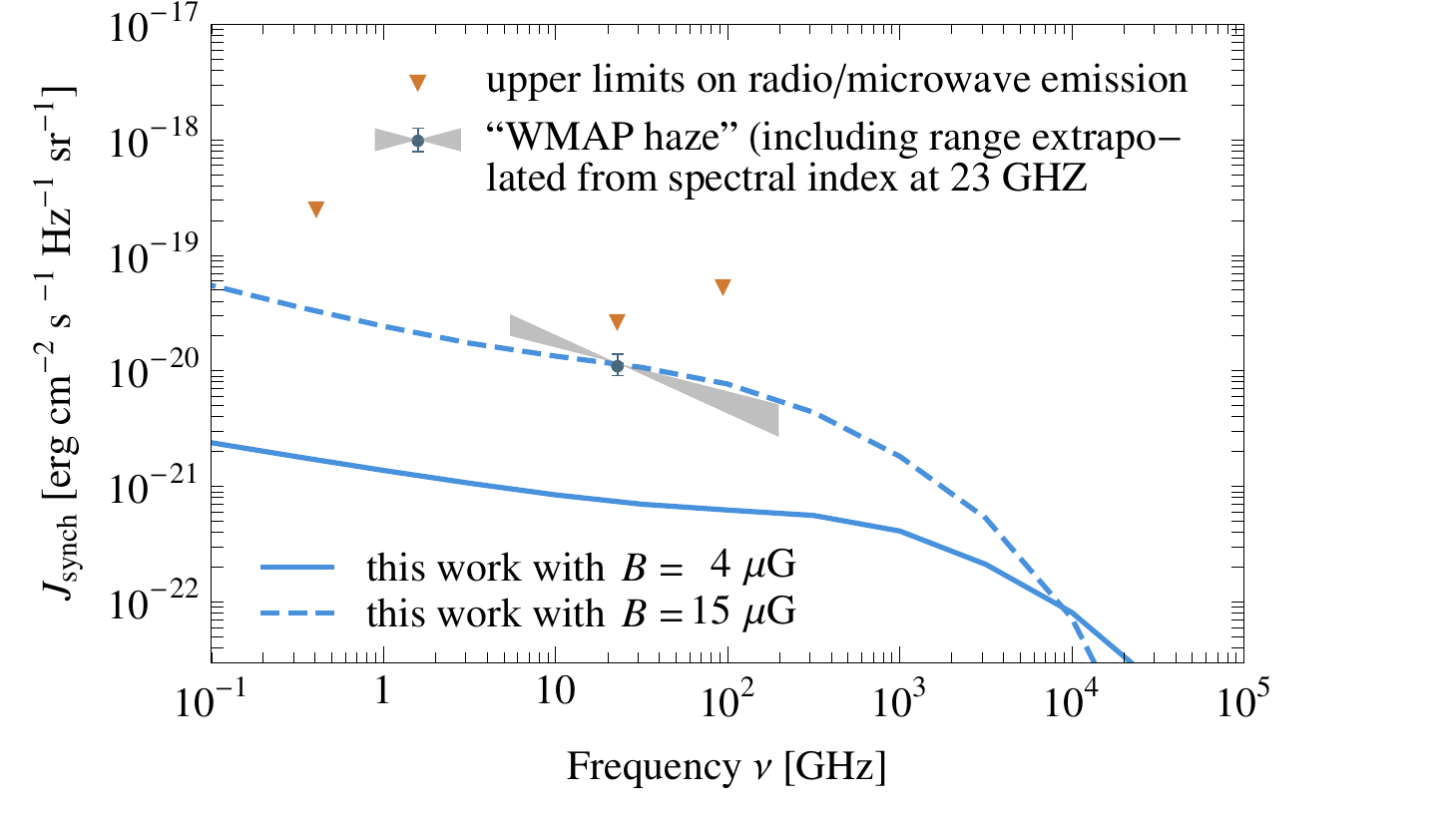}
\end{center}
\caption[]{Our model prediction for the synchrotron flux at $(\ell, b)
  = ( 0^{\circ}, 25^{\circ})$, for two assumed magnetic field values,
  compared to the inferred spectrum of the WMAP
  haze~\cite{Finkbeiner:2004us}.}
\label{fig4}
\end{figure}


We now discuss the parameters that can reproduce the observed
$\gamma$-ray flux --- both its spectrum and morphology.
Kelvin-Helmholtz instabilities have been observed to be generated on
kpc scales in MHD simulations of the Fermi bubble
gas~\cite{Guo:2011eg}, so we choose the scale of turbulence generation
to be $L = 2 \,\text{kpc}$. The shock velocity can in principle be
determined kinematically from the variation of its position with time
(the shock needs
$\sim\!50\,(U/10^8\,\text{cm}\,\text{s}^{-1})\,\text{yr}$ to move a
distance corresponding to the $1''$ resolution of the Chandra X-ray
observatory) or possibly inferred from the observed shock heating. We
fix $U = 2.6 \times 10^8 \,\text{cm}\,\text{s}^{-1}$, a value
consistent with MHD simulations~\cite{Guo:2011eg}. Finally the
Alfv\'en velocity is given by the square root of the ratio of magnetic
field energy density to thermal plasma energy density: $\beta_\text{A}
= v_\text{A}/c = \sqrt{U_B/U_\rho}$. Hence $\beta_\text{A} > 2.8
\times 10^{-4}$ for an estimated upper limit on the thermal gas
density $n \lesssim 10^{-2}\,\text{cm}^{-3}$~\cite{Su:2010qj} and a
magnetic field $B = 4 \, \mu \text{G}$ (suggested by radio
observations of the edge-on spiral galaxy NGC 891~\cite{Beck:1979aa}).
We adopt $\beta_\text{A} = 5 \times 10^{-4}$. The gyro radius of
relativistic electrons is then $\sim\!7.5 \times 10^{11}
(B/{4\,\mu\text{G}})^{-1} (E/ \text{GeV})\,\text{cm}$, which is much
smaller than the dissipation length \mbox{$l_\text{d} > 8 \times
  10^{19} (L/{\text{kpc}}) (U/{10^8\,\text{cm}\,\text{s}^{-1}})^{-3}
  (\beta_\text{A}/{10^{-3}})^3\,\text{cm}$} even for $\mathcal{O}(10)
\, \text{TeV}$ electrons, thus confirming the energy-independence of
the acceleration and escape time. With these parameters we find a
total energy in electrons above $100\,\text{MeV}$ of
$\sim10^{51}\,\text{erg}$ which is over five orders of magnitude
smaller than the required energy in protons in the hadronic emission
model~\cite{Crocker:2010dg}.

In Fig.~\ref{fig2} we show our predicted flux $E^2 J_{\gamma}$ of high
energy $\gamma$-rays (averaged over the surface of the Fermi bubbles)
as a function of energy, and compare it to the data~\cite{Su:2010qj}
as well as to the hadronic~\cite{Crocker:2010dg} and leptonic
DSA~\cite{Cheng:2011xd} models. Note that our hard electron spectrum
nicely reproduces the spectral shoulder around a GeV and the cut-off
at a few hundred GeV.  We show how the total gamma-ray flux is made up
of contributions from IC scattering on the CMB, FIR and optical/UV
backgrounds (dot-dashed line, from left to right). Since we expect the
FIR and optical/UV contributions to decrease rapidly with distance
from the disk, the emission from high latitudes should cut off above
tens of GeV --- a potential test of the model. In Fig.~\ref{fig3} we
compare the data with the predicted $\gamma$-ray intensity at $2$ and
$10 \, \text{GeV}$ as a function of distance from the bubble edge
(calculated as in Ref.~\cite{Su:2010qj}, \emph{i.e.} averaging over
arcs of great circles converging at the bubble centre). This test has
not been done before; it is seen that our model matches the almost
constant intensity in the interior with hard edges, in contrast to the
hadronic~\cite{Crocker:2010dg} and leptonic DSA~\cite{Cheng:2011xd}
models. We also show our model prediction for higher
($500\,\text{GeV}$) $\gamma$-ray energies; the presence of sharp edges
in $\gamma$-rays can be tested by the forthcoming Cherenkov Telescope
Array.


The energy dependence of the profiles reflects the spatial variation
of the electron spectrum with distance from the shock (see also
Fig.~\ref{fig1}). Close to the shock, the acceleration time is
relatively small such that the spectrum is very hard, $p_{\text{eq}}$
is large, and the spectral bump and cut-off appear at high
energies. Further away from the shock, $t_\text{acc}$ becomes larger
and the spectrum softer, while the bump and cut-off move to lower
energies. The emission of the highest energy $\gamma$-rays (due to the
highest energy electrons) is thus localised close to the shock and
results in the limb-brightening above a few hundred GeV. Intermediate
energy $\gamma$-rays can be produced from both high and intermediate
energy electrons which have a more extended distribution, leading to a
flatter intensity profile.

While the `WMAP haze' \cite{Finkbeiner:2004us} has not been observed
in polarised emission~\cite{Gold:2010fm} and may just be an artefact
of the template subtraction~\cite{Mertsch:2010ga}, it has been
proposed as a physical counterpart of the Fermi
bubbles~\cite{Su:2010qj}. However as seen in Fig.~\ref{fig4}, the
expected synchrotron flux in our model is of the required amplitude
only if the magnetic field is as strong as $15\,\mu \text{G}$, several
kpc from the plane.

The hadronic model predicts a detectable flux of neutrinos for the
proposed Mediterranean km$^3$ neutrino
telescope~\cite{Crocker:2010dg}. However the observed bubble profile
disfavours this model (as well as the leptonic DSA model) and instead
favours 2$^\text{nd}$-order Fermi acceleration of electrons, which
would not generate any neutrinos.

\end{document}